\documentstyle[11pt,newpasp,twoside,psfig]{article}
\begin{document}

\title{How relevant is the torus activity/geometry  for the TeV gamma-rays emitted in  the jet of M87 ?}
\author{A.-C. Donea, R. J.  Protheroe}
\affil{Department of Physics and Mathematical Physics, The University of Adelaide, Adelaide, SA 5005, Australia}

\setcounter{page}{111}
\index{de Gaulle, C.}
\index{Churchill, W.}

\begin{abstract}
 Motivated by unification schemes of active galactic nuclei, we review
evidence for the existence of a small-scale dust torus in M87, a
Fanaroff-Riley Class I radio galaxy.  Since there is no direct
evidence of any thermal emission from its torus we consider indirect
evidence, such as BLR activity and  ageing arguments to model the
cold dust structure of M87.  In the context of the jet --
accretion disk -- torus symbiosis we discuss the interactions of GeV
and TeV gamma-rays produced in the jet of M87 with the infrared
radiation fields external to the jet, produced by a less active torus.
A thin and cold torus with less defined outer boundaries could still 
posses problems to some of the TeV emission from the jet.

\end{abstract}

\section{Introduction}

We have already discussed in Donea \& Protheroe (2002) some
aspects related to the IR radiation from ubiquitous dusty tori in
quasars and some blazars.  We have shown that the torus radiation
could present a serious problem for the escape of TeV
$\gamma$-rays from emission regions which are not well above hot
tori.

\begin{figure} [t]
\centerline{\psfig{file=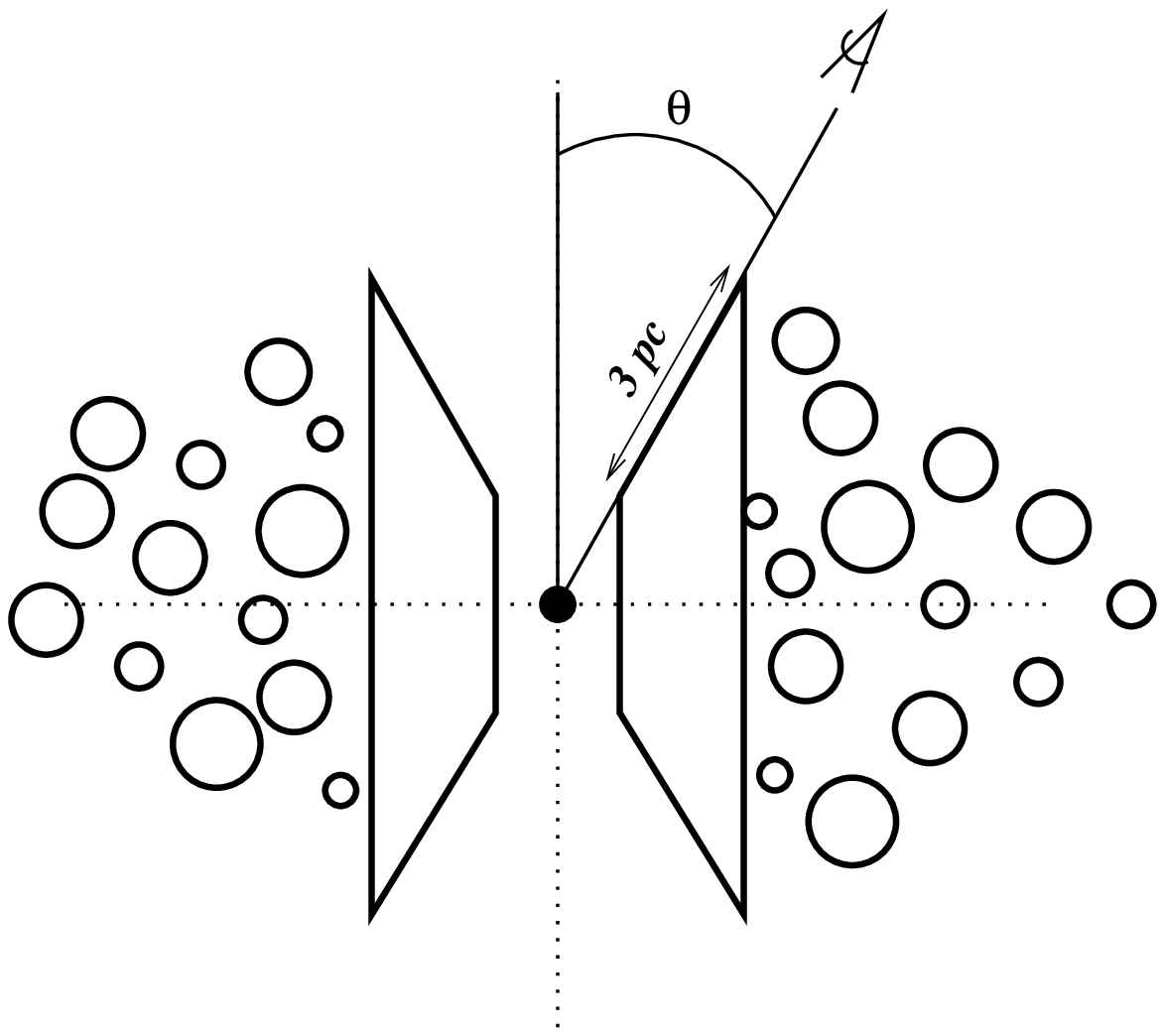,height=6cm,width=7cm}\psfig{file=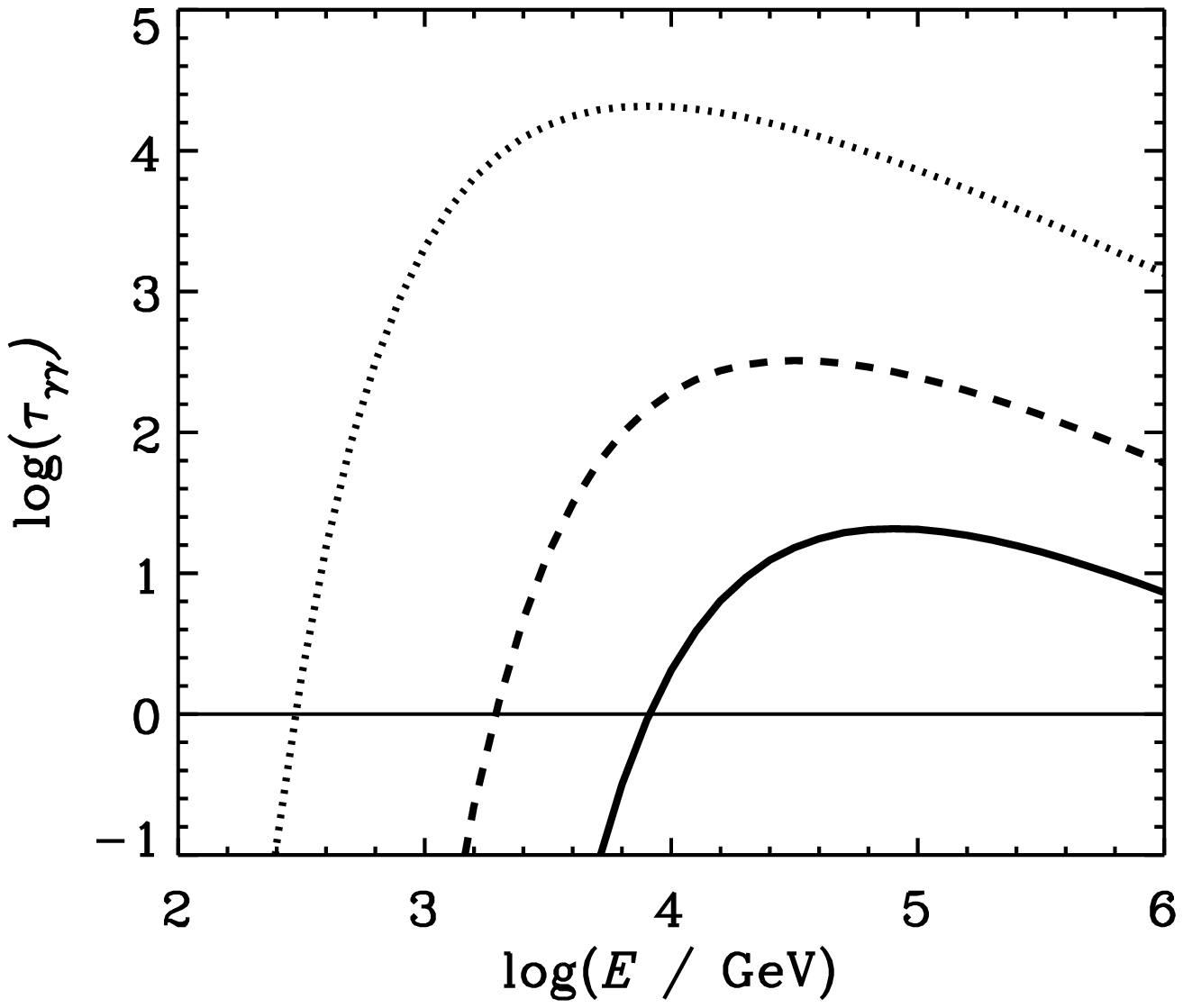,height=6cm,width=7cm}} 
\caption{a) The main body of the torus of M87 is assumed to be cold with $R_{\rm
in}\approx 1$~pc., $R_{\rm out} \approx 2.5$~pc, and its cross section
being trapezoidal with its inner edges cut away at angle
$\phi=30^{\circ}$. For $R>R_{\rm out}$ the dust is highly
inhomogeneous ending up in an diffuse dust-boundary. b) Optical
depth for the absorption of GeV--TeV photons travelling along the jet
from the central 0.001 pc, in the weak infrared bath of photons from
this torus. The  curves are calculated for different temperatures
of the dust $T_0=(100, 250, 1000)$~K (from bottom to top).
\label{Fig1}}
\end{figure}


It is hard to accept that M87, a mis-aligned BL Lac object, that
posses a huge black hole (Ford et al., 1995), a very active jet
(Wilson et al., 2002) and BLR (Sankrit et al., 1999) and possibly NLR
(Zavala \& Taylor, 2002), completely lacks the dusty component, be it
very cold.  M87 has a low luminosity nucleus and the fact that strong
mid-infrared emission was not detected from the centre of M87 (Perlman
et al., 2001) suggests either that there is no torus in M87, or if
there is one it should have a low radiative efficiency making its
detection difficult.  Using argument related to the age of M87, the
fact that M87 could undergo on-off activity cycles (Di Matteo et al.,
2002) and the possibility that M87 could harbour a double black hole
system (Zier \& Biermann, 2001) we assume that M87 could still
have sufficient dust around its nucleus (Fig. 1a).

We calculate the optical depth for $\gamma$-$\gamma$ pair-production
$\tau_{\gamma\gamma}$ of GeV-TeV photons emitted by the jet
interacting with IR photons from the torus (Fig. 1b).  Bai and Lee
(2002) have suggested that M87 could emit detectable TeV photons with
an inverse-Compton emission peak at only 0.1 TeV.  Protheroe et
al. (2002) have predicted that $\gamma$-ray emission involving
hadronic interactions cuts off at 10 GeV - 1 TeV, depending on the
luminosity of the low hump of SED. We find that $\gamma$-rays with
energies around 0.1 TeV are not absorbed by the IR cold photons and
therefore could be detectable. Although the high temperature of
$1000$~K is unrealistic for the present near-dormant state of M87, it
is interesting to see what the difference would be in absorption when
M87 would power on the torus.

We note that $\gamma$-rays with energies above 10 TeV (if produced)
could be absorbed even when the torus is cold ($T=100$~K) and thin,
and the emission region lies within the torus. As one moves away from
the black hole, the TeV emission if produced in the jet, escapes the
$\gamma$-$\gamma$ absorption and could be detected by the Cherenkov
telescopes.  We conclude that the absorption of GeV to 10 TeV photons
by $\gamma$-$\gamma$ interactions with the infrared photons produced
by a cold torus is negligible.  Only a strong and rapid
ignition in the  nucleus of M87 could heat up the dust and 
increase the absorption of $\gamma$-rays produced nearby the infrared torus.


\vspace{-0.5cm}
 
\begin{references}
\vspace{-0.3cm}
\reference Bai J. M., Lee M. G., 2001, ApJ, 549, L173.
\reference Di Matteo T., et al., astro-ph 0202238
\reference A.-C. Donea, R.J. Protheroe, 2002, Astroph. Phys, to appear.
\reference R.J. Protheroe, A.-C. Donea, Reimer A,  2002, Astroph. Phys, submitted.
\reference Perlman, E. S., et al., 2001, ApJ, 561, L51.
\reference Sankrit, R., Sembach, K. R., \& Canizares, C. R. 1999, ApJ, 527, 733
\reference Zavala R. T., Taylor G. B., 2002, ApJ, 566, 9.
\reference Zier C., Biermann P. L., 2001, A\&A, 377, 23.
\reference Wilson A.S., Yang, Y., ApJ, 568,133, 2002.
\end{references}
\end{document}